\documentstyle[prl,floats,multicol,epsf,psfrag,amssymb,amsgen,
amsfonts,amsbsy,aps]{revtex}

\tighten
\begin{document}
\draft
\title{Greenberger-Horne-Zeilinger nonlocality in phase space}
\author{P.\ van Loock and Samuel L.\ Braunstein}
\address{Quantum Optics and Information Group,\\
School of Informatics, University of Wales, Bangor LL57 1UT, UK}
\maketitle

\begin{abstract}
We show that the continuous-variable analogues to the multipartite entangled
Greenberger-Horne-Zeilinger states of qubits violate Bell-type inequalities
imposed by local realistic theories. Our results suggest that 
the degree of nonlocality of these nonmaximally 
entangled continuous-variable states, represented by the maximum violation, 
grows with increasing number of parties. This growth does not appear to be 
exponentially large as for the maximally entangled qubit states, but rather
decreases for larger numbers of parties.
\end{abstract}
\pacs{PACS numbers: 03.67.-a, 03.65.Bz, 42.50.Dv}
\vspace{3ex}

\begin{multicols}{2}

Entanglement and nonlocality are the most outstanding 
features of quantum mechanics. In the rapidly advancing field of
quantum communication and computation, entangled states
are the key ingredients: they enable quantum teleportation \cite{Benn},
quantum cryptography \cite{Ekert}, and many other potentially useful schemes. 
Bell showed that nonlocality can be revealed via the constraints that
local realism imposes on the statistics of two physically separated 
systems \cite{Bell}. These constraints, expressed in
terms of the Bell inequalities, can be violated by quantum mechanics.
Entanglement does not automatically imply nonlocality.
The so-called Werner states are mixed states which are  
inseparable, but do not violate any Bell inequality \cite{Werner}.
Also pure entangled states can, if associated
with a positive Wigner function, directly reveal a local
hidden-variable description \cite{Bell}.

Towards possible applications in quantum communication, both theoretical 
and experimental investigations increasingly focus on quantum states
with a continuous spectrum defined in an infinite dimensional Hilbert
space. These states can be relatively easily generated using squeezed light
and beam splitters, as for instance the entangled two-mode squeezed 
vacuum state that has already proven its usefulness for quantum 
teleportation \cite{Furu}. The two-mode squeezed vacuum state is an
approximate version of the original Einstein-Podolsky-Rosen (EPR) state
\cite{EPR} where the quadrature amplitudes of the electromagnetic field
play the roles of position and momentum of a particle. Its Wigner function 
is positive everywhere and hence it has a local hidden-variable
description \cite{Bell}. 
Thus, attempts to derive for this state violations of Bell 
inequalities based on homodyne measurements of the quadratures 
failed \cite{Ou}.
However, whether nonlocality is uncovered depends on the observables
and the measurements considered in a specific Bell inequality and not only
on the quantum state itself.
It was shown by Banaszek and Wodkiewicz \cite{Bana}, 
that the two-mode squeezed vacuum state is nonlocal, as it violates
a Clauser-Horne-Shimony-Holt (CHSH) inequality \cite{CHSH} when
measurements of photon number parity are considered.

The nonlocality of the multipartite entangled Green\-berger-Horne-Zeilinger 
(GHZ) states can {\it in principle} be manifest in a single measurement 
and need not be statistical \cite{GHZ} as the violation of a Bell 
inequality that relies on mean values.
But Mermin and others \cite{Mermin,Gisin} also derived Bell-CHSH inequalities
for $N$-particle systems. The aim of this paper is to apply those
$N$-party inequalities to continuous-variable 
GHZ states \cite{PvL} and thereby to prove their nonlocality. 
Since these states have a positive Wigner function, 
we shall follow the convenient strategy of Banaszek and Wodkiewicz 
\cite{Bana} who exploited the fact that the Wigner function is connected 
to the quantum mean value of the photon number parity operator. 
Relying on this connection, we will demonstrate $N$-party nonlocality 
using mean-value inequalities \cite{Gisin}, and we do not 
follow the original GHZ program utilizing a contradiction to local realism 
in a single measurement.

Let us identify the `position' and `momentum' of a particle 
with the quadrature amplitudes of a single electromagnetic mode 
(the real and imaginary part of the mode's annihilation
operator). In Ref.~\onlinecite{PvL}, it has been shown that a sequence
of beam splitter operations,
$\hat{B}_{N-1\,N}(\pi/4)\hat{B}_{N-2\,N-1}
\left(\cos^{-1}1/\sqrt{3}\right)
\times\cdots\times\hat{B}_{12}\left(\cos^{-1}1/\sqrt{N}\right)$,
applied to one momentum squeezed vacuum mode 1 and $N-1$ position
squeezed vacuum modes $2$ through $N$, yields an $N$-mode state
with $N$-party entanglement between all modes. Here, an ideal (phase-free) 
beamsplitter operation $\hat{B}_{ij}(\theta)$ acts on a pair of modes 
$i$ and $j$ with annihilation operators $\hat a_i$ and $\hat a_j$ like  
$\hat a_i \rightarrow \hat a_i \cos\theta + \hat a_j\sin\theta$,
$\hat a_j \rightarrow \hat a_i \sin\theta - \hat a_j\cos\theta$.
The Wigner function of the pure entangled $N$-mode state is
\end{multicols}
\noindent\rule{5cm}{.6pt}
\begin{eqnarray}\label{Wignerxp}
W({\bf x},{\bf p})
=\left(\frac{2}{\pi}\right)^N
\exp\left\{-e^{-2r}\left[\frac{2}{N}\left(\sum_{i=1}^N x_i\right)^2+
\frac{1}{N}\sum_{i,j}^N(p_i-p_j)^2\right]
-e^{+2r}\left[\frac{2}{N}\left(\sum_{i=1}^N p_i\right)^2
+\frac{1}{N}\sum_{i,j}^N(x_i-x_j)^2\right]\right\} \;,
\end{eqnarray}
\begin{multicols}{2}
\noindent where ${\bf x}=x_1,x_2,...,x_N$ and ${\bf p}=p_1,p_2,...,p_N$
are the positions and momenta of the $N$ modes and $r$ is the squeezing
parameter (with equal squeezing in all initial modes).
The state $W({\bf x},{\bf p})$ is always positive, symmetric among the 
$N$ modes, and becomes peaked at $x_i-x_j=0$ ($i,j=1,2,...,N$) and 
$p_1+p_2+\cdots+p_N=0$ for large squeezing $r$. 
For $N=2$, it equals the well-known EPR-state Wigner function which 
approaches $\delta(x_1-x_2)\delta(p_1+p_2)$ in the limit of infinite 
squeezing. Any nonzero squeezing yields $N$-partite entanglement 
and the position and momentum correlations can be exploited for 
quantum teleportation \cite{Sam98a} between any two of $N$ parties 
with the assistance of the remaining $N-2$ parties \cite{PvL}. 

In order to prove the nonlocality exhibited by the state 
$W({\bf x},{\bf p})$, we use the fact that the Wigner function 
is proportional to the quantum expectation value of a displaced parity 
operator \cite{Royer}. 
We obtain the relation \cite{Bana}
\begin{eqnarray}\label{parity1}
W({\boldsymbol{\alpha}})=\left(\frac{2}{\pi}\right)^N\left\langle
\hat{\Pi}({\boldsymbol{\alpha}})\right\rangle=
\left(\frac{2}{\pi}\right)^N\Pi({\boldsymbol{\alpha}}) \;,
\end{eqnarray}
where ${\boldsymbol{\alpha}}={\bf x}+i{\bf p}=\alpha_1,\alpha_2,...,\alpha_N$
and $\Pi({\boldsymbol{\alpha}})$ is the quantum expectation value of the
operator
\begin{eqnarray}\label{parity2}
\hat{\Pi}({\boldsymbol{\alpha}})=\bigotimes_{i=1}^N\hat{\Pi}_i(\alpha_i)=
\bigotimes_{i=1}^N\hat{D}_i(\alpha_i)
(-1)^{\hat{n}_i}\hat{D}_i^{\dagger}(\alpha_i) \;.
\end{eqnarray}
The operators $\hat{D}_i(\alpha_i)$ are phase-space displacement
operators acting on mode $i$.
Thus, $\hat{\Pi}({\boldsymbol{\alpha}})$ is a product of displaced parity 
operators given by
\begin{eqnarray}\label{parity3}
\hat{\Pi}_i(\alpha_i)=\hat{\Pi}_i^{(+)}(\alpha_i)-
\hat{\Pi}_i^{(-)}(\alpha_i) \;,
\end{eqnarray}
with the projection operators
\begin{eqnarray}\label{parity4}
\hat{\Pi}_i^{(+)}(\alpha_i)&=&\hat{D}_i(\alpha_i)\sum_{k=0}^{\infty}
|2k\rangle\langle 2k|\hat{D}_i^{\dagger}(\alpha_i),\\
\hat{\Pi}_i^{(-)}(\alpha_i)&=&\hat{D}_i(\alpha_i)\sum_{k=0}^{\infty}
|2k+1\rangle\langle 2k+1|\hat{D}_i^{\dagger}(\alpha_i) \;,
\end{eqnarray}
corresponding to the measurement of an even (parity $+1$) or an odd 
(parity $-1$) number of photons in mode $i$. 
This means that each mode is now characterized
by a dichotomic variable similar to the single-particle spin or the 
single-photon polarization. Different spin or polarizer orientations
are replaced by different displacements in phase space.
These different settings of a measurement with two
possible outcomes $\pm 1$ for each possible setting is exactly what
we need for the nonlocality test.

In the case of $N$-particle systems, such a nonlocality test is possible
using the $N$-particle generalization of the two-particle Bell-CHSH
inequality \cite{Gisin}. This inequality is based on the following
recursively defined linear combination of joint measurement results
\begin{eqnarray}\label{lincomb1}
B_N&\equiv&\case{1}{2}[\sigma(a_N)+\sigma(a_N')]B_{N-1}\nonumber\\
&+&\case{1}{2}[\sigma(a_N)-\sigma(a_N')]B_{N-1}'=\pm 2 \;,
\end{eqnarray}
where $\sigma(a_N)=\pm 1$ and $\sigma(a_N')=\pm 1$ describe two possible 
outcomes for two possible measurement settings (denoted by $a_N$ and
$a_N'$) of measurements on the $N$th particle.
Provided that $B_{N-1}=\pm 2$ and $B_{N-1}'=\pm 2$, Equation (\ref{lincomb1}) 
is true for a single run of the measurements
where $\sigma(a_N)$ becomes either $+1$ or $-1$ and so does $\sigma(a_N')$.
Thus, induction proves Eq.~(\ref{lincomb1}) for any $N$ with
\begin{eqnarray}\label{lincomb2}
B_2&\equiv& [\sigma(a_1)+\sigma(a_1')]\sigma(a_2)\nonumber\\
&+&[\sigma(a_1)-\sigma(a_1')]\sigma(a_2')=\pm 2 \;,
\end{eqnarray}
which is trivially true (the expressions $B_{N}'$ are equivalent to
$B_{N}$ but with all the $a_i$ and $a_i'$ swapped).
Within the framework of local realistic theories with the hidden variables
${\boldsymbol{\lambda}}=\lambda_1,\lambda_2,...,\lambda_N$ and the normalized 
probability distribution $P({\boldsymbol{\lambda}})$, we obtain an inequality
for the average value of $B_N\equiv B_N({\boldsymbol{\lambda}})$,
\begin{eqnarray}\label{ineq1}
\left|\int d\lambda_1 d\lambda_2 ... d\lambda_N P({\boldsymbol{\lambda}})
B_N({\boldsymbol{\lambda}})\right| \leq 2 \;.
\end{eqnarray}
By the linearity of averaging, this is a sum of means of products
of the $\sigma(a_i)$ and $\sigma(a_i')$. For example, if $N=2$, we obtain 
the CHSH inequality
\begin{eqnarray}\label{ineq2}
|C(a_1,a_2)+C(a_1,a_2')+C(a_1',a_2)-C(a_1',a_2')| \leq 2,
\end{eqnarray}
with the correlation functions 
\begin{eqnarray}\label{corrfct2}
C(a_1,a_2)=\int d\lambda_1 d\lambda_2 P(\lambda_1,\lambda_2) 
\sigma(a_1,\lambda_1)\sigma(a_2,\lambda_2) \;.
\end{eqnarray}
Following Bell \cite{Bell}, an always positive Wigner
function can serve as the hidden-variable probability distribution.
In this sense, the EPR-state Wigner function could prevent the CHSH
inequality being violated: $W(x_1,p_1,x_2,p_2)\equiv 
P(\lambda_1,\lambda_2)$. The same applies to the general Wigner 
function in Eq.~(\ref{Wignerxp}): $W({\bf x},{\bf p})\equiv 
P({\boldsymbol{\lambda}})$
could be used to construct correlation functions 
\begin{eqnarray}\label{corrfctN}
C({\bf a})&=&\int d\lambda_1 d\lambda_2 ... d\lambda_N 
P({\boldsymbol{\lambda}})\nonumber\\
&&\times\;\sigma(a_1,\lambda_1)\sigma(a_2,\lambda_2)\cdots 
\sigma(a_N,\lambda_N),
\end{eqnarray} 
where ${\bf a}=a_1,a_2,...,a_N$.
However, for parity measurements on each mode with possible results
$\pm 1$ and different settings by different displacements, this would 
require unbounded $\delta$ functions for the local objective quantities
$\sigma(a_i,\lambda_i)$ \cite{Bana}, as in this case the relation
\begin{eqnarray}\label{relation}
C({\bf a})\equiv\Pi({\boldsymbol{\alpha}})=
(\pi/2)^N W({\boldsymbol{\alpha}})
\end{eqnarray} 
holds.
This relation, which directly relates the correlation function to the 
Wigner function, is indeed crucial for the nonlocality proof of the 
continuous-variable states in Eq.~(\ref{Wignerxp}). 
For the EPR state with $N=2$, we can now look at the combination \cite{Bana}
\begin{eqnarray}\label{B2}
{\mathcal{B}}_2=\Pi(0,0)+\Pi(0,\beta)+\Pi(\alpha,0)-\Pi(\alpha,\beta) \;,
\end{eqnarray}
which according to Eq.~(\ref{ineq2}) satisfies $|{\mathcal{B}}_2|\leq 2$
for local realistic theories.
Here, we have chosen the displacement settings $\alpha_1=\alpha_2=0$
and $\alpha_1'=\alpha$, $\alpha_2'=\beta$.

Let us write the states in Eq.~(\ref{Wignerxp}) as
\end{multicols}
\begin{eqnarray}\label{Wigneralpha}
\Pi({\boldsymbol{\alpha}})=
\exp\left\{-2\cosh 2r\sum_{i=1}^N |\alpha_i|^2+
\sinh 2r\left[\frac{2}{N}\sum_{i,j}^N(\alpha_i\alpha_j
+\alpha_i^*\alpha_j^*)-\sum_{i=1}^N
(\alpha_i^2+\alpha_i^{*2})\right]\right\} \;.
\end{eqnarray}
\hfill\rule{5cm}{.6pt}
\begin{multicols}{2}
For $N=2$ and $\alpha=\beta=i\sqrt{\mathcal{J}}$ in terms of
the real displacement parameter ${\mathcal{J}}\geq 0$ \cite{note},
these states yield ${\mathcal{B}}_2=1+2\exp(-2{\mathcal{J}}\cosh 2r)
-\exp(-4{\mathcal{J}}e^{+2r})$. In the limit of large $r$ 
($\cosh 2r\approx e^{+2r}/2$) and small 
${\mathcal{J}}$, this ${\mathcal{B}}_2$ is maximized for 
${\mathcal{J}}e^{+2r}=(\ln 2)/3$: ${\mathcal{B}}_2^{\rm max}\approx 2.19$
\cite{Bana}, which is a clear violation of the inequality
$|{\mathcal{B}}_2|\leq 2$. Smaller violations occur also for smaller
squeezing and bigger ${\mathcal{J}}$. For any nonzero squeezing, some
violation takes place (see Fig.~1).

Let us now examine the three-mode state and set $N=3$ in 
Eq.~(\ref{Wigneralpha}). According to the inequality of the correlation
functions derived from Eq.~(\ref{lincomb1})-(\ref{ineq1}) with $N=3$,
\begin{eqnarray}\label{ineq3}
|&&C(a_1,a_2,a_3')+C(a_1,a_2',a_3)\nonumber\\
&&+C(a_1',a_2,a_3)-C(a_1',a_2',a_3')| \leq 2 \;,
\end{eqnarray} 
for the possible combination
\begin{eqnarray}\label{B3}
{\mathcal{B}}_3=\Pi(0,0,\gamma)+\Pi(0,\beta,0)+\Pi(\alpha,0,0)
-\Pi(\alpha,\beta,\gamma),
\end{eqnarray}
a contradiction to local realism does not occur only if
$|{\mathcal{B}}_3|\leq 2$. The corresponding settings here are
$\alpha_1=\alpha_2=\alpha_3=0$ and $\alpha_1'=\alpha$, $\alpha_2'=\beta$,
$\alpha_3'=\gamma$.
With the choice $\alpha=\sqrt{\mathcal{J}}e^{i\phi_1}$,
$\beta=\sqrt{\mathcal{J}}e^{i\phi_2}$, and 
$\gamma=\sqrt{\mathcal{J}}e^{i\phi_3}$, we obtain 
\end{multicols}
\noindent\rule{5cm}{.6pt}
\begin{eqnarray}\label{B3long}
{\mathcal{B}}_3=\sum_{i=1}^3\exp(-2{\mathcal{J}}\cosh 2r-\case{2}{3}
{\mathcal{J}}\sinh 2r\cos 2\phi_i)
-\exp\left\{-6{\mathcal{J}}\cosh 2r-\case{1}{3}{\mathcal{J}}\sinh 2r
\sum_{i\neq j}^3[\cos 2\phi_i-4\cos(\phi_i+\phi_j)]\right\}.
\end{eqnarray}
\hfill\rule{5cm}{.6pt}
\begin{multicols}{2}
Apparently, because of the symmetry of the entangled three-mode state,
equal phases $\phi_i$ should also be chosen in order to maximize
${\mathcal{B}}_3$. The best choice is $\phi_1=\phi_2=\phi_3=\pi/2$,
which ensures that the positive terms in Eq.~(\ref{B3long})
become maximal and the contribution of the negative term minimal.
Therefore we again use equal settings $\alpha=\beta=\gamma=
i\sqrt{\mathcal{J}}$ and obtain
\begin{eqnarray}\label{B3choice}
{\mathcal{B}}_3&=&3\exp(-2{\mathcal{J}}\cosh 2r+2{\mathcal{J}}\sinh 2r/3)
\nonumber\\
&-&\exp(-6{\mathcal{J}}e^{+2r}) \;.
\end{eqnarray} 
The violations of $|{\mathcal{B}}_3|\leq 2$ that occur with this result
are similar to the violations $|{\mathcal{B}}_2|\leq 2$ obtained for 
the EPR state, but the $N=3$ violations 
are even more significant than the $N=2$ violations (see Fig.~1).
In the limit of large $r$ (and small ${\mathcal{J}}$), 
we may use $\cosh 2r\approx\sinh 2r\approx
e^{+2r}/2$ in Eq.~(\ref{B3choice}). Then ${\mathcal{B}}_3$ is maximized for 
${\mathcal{J}}e^{+2r}=3(\ln 3)/16$: ${\mathcal{B}}_3^{\rm max}\approx 2.32$.
This requires even smaller displacements ${\mathcal{J}}$ than in the
$N=2$ case for the same squeezing.

Let us now investigate the cases $N=4$ and $N=5$.
From Eq.~(\ref{lincomb1})-(\ref{ineq1}) with $N=4$, the following
inequality for the correlation functions can be derived:
\begin{eqnarray}\label{ineq4}
\case{1}{2}|&&C(a_1,a_2,a_3,a_4')+C(a_1,a_2,a_3',a_4)+C(a_1,a_2',a_3,a_4)
\nonumber\\
+&&C(a_1',a_2,a_3,a_4)+C(a_1,a_2,a_3',a_4')+C(a_1,a_2',a_3,a_4')\nonumber\\
+&&C(a_1',a_2,a_3,a_4')+C(a_1,a_2',a_3',a_4)+C(a_1',a_2,a_3',a_4)\nonumber\\
+&&C(a_1',a_2',a_3,a_4)-C(a_1',a_2',a_3',a_4)-C(a_1',a_2',a_3,a_4')
\nonumber\\
-&&C(a_1',a_2,a_3',a_4')-C(a_1,a_2',a_3',a_4')-C(a_1,a_2,a_3,a_4)
\nonumber\\
-&&C(a_1',a_2',a_3',a_4')| \leq 2 \;.
\end{eqnarray} 
It is symmetric among all four parties as any inequality derived 
from Eq.~(\ref{lincomb1})-(\ref{ineq1}) is symmetric among all parties.
For the settings $\alpha_1=\alpha_2=\alpha_3=\alpha_4=0$
and $\alpha_1'=\alpha$, $\alpha_2'=\beta$, $\alpha_3'=\gamma$,
$\alpha_4'=\delta$, complying with local realism implies 
$|{\mathcal{B}}_4|\leq 2$ where
\begin{eqnarray}\label{B4}
{\mathcal{B}}_4=\case{1}{2}[&&\Pi(0,0,0,\delta)+\Pi(0,0,\gamma,0)
+\Pi(0,\beta,0,0)\nonumber\\
+&&\Pi(\alpha,0,0,0)+\Pi(0,0,\gamma,\delta)+\Pi(0,\beta,0,\delta)
\nonumber\\
+&&\Pi(\alpha,0,0,\delta)+\Pi(0,\beta,\gamma,0)+\Pi(\alpha,0,\gamma,0)
\nonumber\\
+&&\Pi(\alpha,\beta,0,0)-\Pi(\alpha,\beta,\gamma,0)
-\Pi(\alpha,\beta,0,\delta)\nonumber\\
-&&\Pi(\alpha,0,\gamma,\delta)-\Pi(0,\beta,\gamma,\delta)
-\Pi(0,0,0,0)\nonumber\\
-&&\Pi(\alpha,\beta,\gamma,\delta)] \;. 
\end{eqnarray}
Similarly, for $N=5$ one finds
\begin{eqnarray}\label{B5}
{\mathcal{B}}_5=\case{1}{2}[&&\Pi(0,0,0,\delta,\epsilon)
+\Pi(0,0,\gamma,0,\epsilon)+\Pi(0,\beta,0,0,\epsilon)\nonumber\\
+&&\Pi(\alpha,0,0,0,\epsilon)+\Pi(0,0,\gamma,\delta,0)
+\Pi(0,\beta,0,\delta,0)\nonumber\\
+&&\Pi(\alpha,0,0,\delta,0)+\Pi(0,\beta,\gamma,0,0)
+\Pi(\alpha,0,\gamma,0,0)\nonumber\\
+&&\Pi(\alpha,\beta,0,0,0)-\Pi(\alpha,\beta,\gamma,\delta,0)
-\Pi(\alpha,\beta,\gamma,0,\epsilon)\nonumber\\
-&&\Pi(\alpha,\beta,0,\delta,\epsilon)-\Pi(\alpha,0,\gamma,\delta,\epsilon)
-\Pi(0,\beta,\gamma,\delta,\epsilon)\nonumber\\
-&&\Pi(0,0,0,0,0)] \;, 
\end{eqnarray}
which has to statisfy $|{\mathcal{B}}_5|\leq 2$ and contains the same
settings as for $N=4$, but in addition $\alpha_5=0$ and 
$\alpha_5'=\epsilon$.

We can now use the entangled states in Eq.~(\ref{Wigneralpha})
with $N=4$ and $N=5$ and apply the inequalities to them.
For the same reason as for $N=3$ (symmetry among all modes
in the states and in the inequalities), the choice $\alpha=\beta=\gamma=
\delta=\epsilon=i\sqrt{\mathcal{J}}$ appears to be optimal
(maximizes positive terms and minimizes negative contributions). 
\end{multicols}

\begin{figure}[htb]
\begin{center}
\begin{psfrags}
     \psfrag{B2}[cb]{\Large ${\mathcal{B}}_2~~~$}
     \psfrag{B3}[cb]{\Large ${\mathcal{B}}_3~~~$}
     \psfrag{B4}[cb]{\Large ${\mathcal{B}}_4~~~$}
     \psfrag{B5}[cb]{\Large ${\mathcal{B}}_5~~~$}
     \psfrag{r}[c]{\Large ~~~~~~~~$r$}
     \psfrag{J}[c]{\Large ${\mathcal{J}}$}
     \psfrag{N=2}{\Large \bf N=2}
     \psfrag{N=3}{\Large \bf N=3}
     \psfrag{N=4}{\Large \bf N=4}
     \psfrag{N=5}{\Large \bf N=5}
\epsfxsize=6.5in
     \epsfbox[-20 20 400 280]{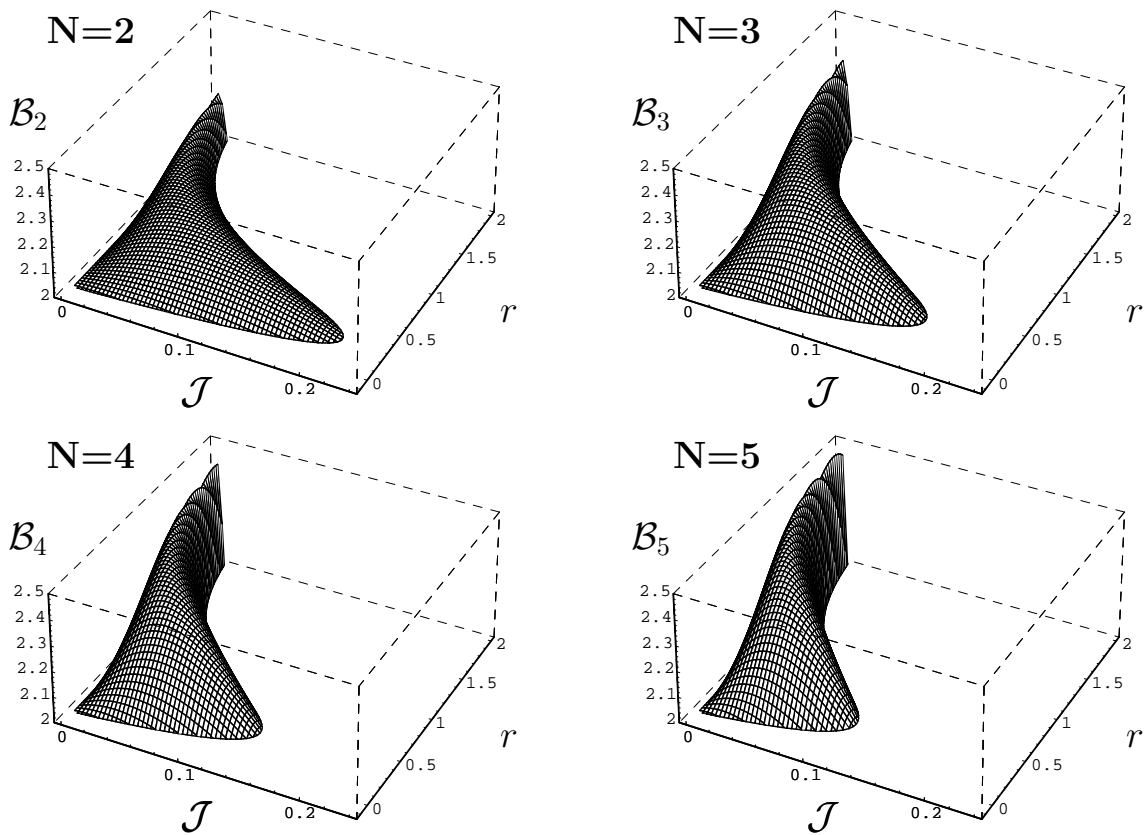}
\end{psfrags}
\end{center}
\caption{Violations of the inequality $|{\mathcal{B}}_N|\leq 2$
imposed by local realistic theories
with the entangled two-mode EPR ($N=2$, as in Ref.~\protect\cite{Bana}), 
three-mode GHZ ($N=3$), four-mode GHZ ($N=4$), and 
five-mode GHZ ($N=5$) states.} 
\label{fig1}
\end{figure}

\begin{multicols}{2}
\noindent With this choice, we obtain
\begin{eqnarray}\label{B4andB5}
{\mathcal{B}}_4&=&2\exp(-2{\mathcal{J}}\cosh 2r+{\mathcal{J}}\sinh 2r)
\nonumber\\
&-&2\exp(-6{\mathcal{J}}\cosh 2r-3{\mathcal{J}}\sinh 2r)\nonumber\\
&+&3\exp(-4{\mathcal{J}}\cosh 2r)
-\case{1}{2}\exp(-8{\mathcal{J}}e^{+2r})-\case{1}{2} \;,\nonumber\\
{\mathcal{B}}_5&=&5\exp(-4{\mathcal{J}}\cosh 2r+4{\mathcal{J}}\sinh 2r/5)
\nonumber\\
&-&\case{5}{2}\exp(-8{\mathcal{J}}\cosh 2r-24{\mathcal{J}}\sinh 2r/5)
-\case{1}{2} \;.
\end{eqnarray}
As shown in Fig.~1, the maximum violation of $|{\mathcal{B}}_N|\leq 2$
(for our particular choice of settings) grows with increasing number
of parties $N$. The asymptotic analysis (large $r$ and small 
${\mathcal{J}}$) yields for $N=5$: ${\mathcal{B}}_5^{\rm max}\approx 2.48$
with ${\mathcal{J}}e^{+2r}=5(\ln 2)/24$.
At a certain amount of large squeezing, smaller displacements 
${\mathcal{J}}$ than for $N\leq 4$ (at the same squeezing) are needed 
to approach this maximum violation. Another important observation
is that in all four cases ($N=2,3,4,5$), violations occur for any nonzero 
squeezing. This requires the presence of $N$-partite entanglement for
any nonzero squeezing which is consistent with the results in
Ref.~\onlinecite{PvL}. Moreover, we see that not only for large
squeezing but also for modest finite squeezing, the significance of the
violations (at optimal displacements ${\mathcal{J}}$) grows with 
increasing $N$. 

In the following, we will examine the general case of $N$ parties.
How does the maximum violation of the Bell-type inequalities
derived with the continuous-variable GHZ states in general
evolve with increasing number of parties, in particular, compared to
the exponential growth for the qubit GHZ states \cite{Mermin,Gisin}?
At least for $N\leq 5$, the maximum violation grows, and this growth 
does not appear to be exponentially
large, but rather seems to decrease. This conjecture has not been proven,
since we did not consider all possible settings (all possible combinations
of $\alpha_i$ and $\alpha_i'$). However, there are strong hints
that our choice of $\alpha_i=0$ and $\alpha_i'=i\sqrt{\mathcal{J}}$
is near optimal. In particular, that the nonlocality is always revealed
for arbitrarily small squeezing (any nonzero squeezing) lets our choice 
appear more appropriate than other possible combinations.
Having now much confidence in the choice of settings that we used for 
small numbers of parties, we will use the same settings for larger 
numbers of parties.

Considering odd numbers of parties $N$, we find the following expression
for ${\mathcal{B}}_N$,
\begin{eqnarray}\label{BN1}
{\rm if}\;N&=&3+8M:\;\;{\mathcal{B}}_N=2^{\frac{3-N}{2}}
\sum_{k=0}^{\frac{N-1}{2}}(-1)^k {N \choose 2k+1}\nonumber\\
&&\times\;\Pi(\alpha_1',\alpha_2',...,\alpha_{2k+1}',
\alpha_{2k+2},\alpha_{2k+3},...,\alpha_N),
\end{eqnarray}
where the first $2k+1$ arguments of $\Pi$ are
$\alpha_1'=\alpha_2'=$
\end{multicols}

\begin{figure}[htb]
\begin{center}
\begin{psfrags}
     \psfrag{B}[cb]{\Large ${\mathcal{B}}_N~~~$}
     \psfrag{A}[c]{\Large ${\mathcal{A}}$}
     \psfrag{N=5}{\Large \bf N=5}
     \psfrag{N=9}{\Large \bf N=9}
     \psfrag{N=15}{\Large \bf N=15}
     \psfrag{N=25}{\Large \bf N=25}
     \psfrag{N=45}{\Large \bf N=45}
     \psfrag{N=85}{\Large \bf N=85}
\epsfxsize=6.5in
     \epsfbox[-20 20 450 230]{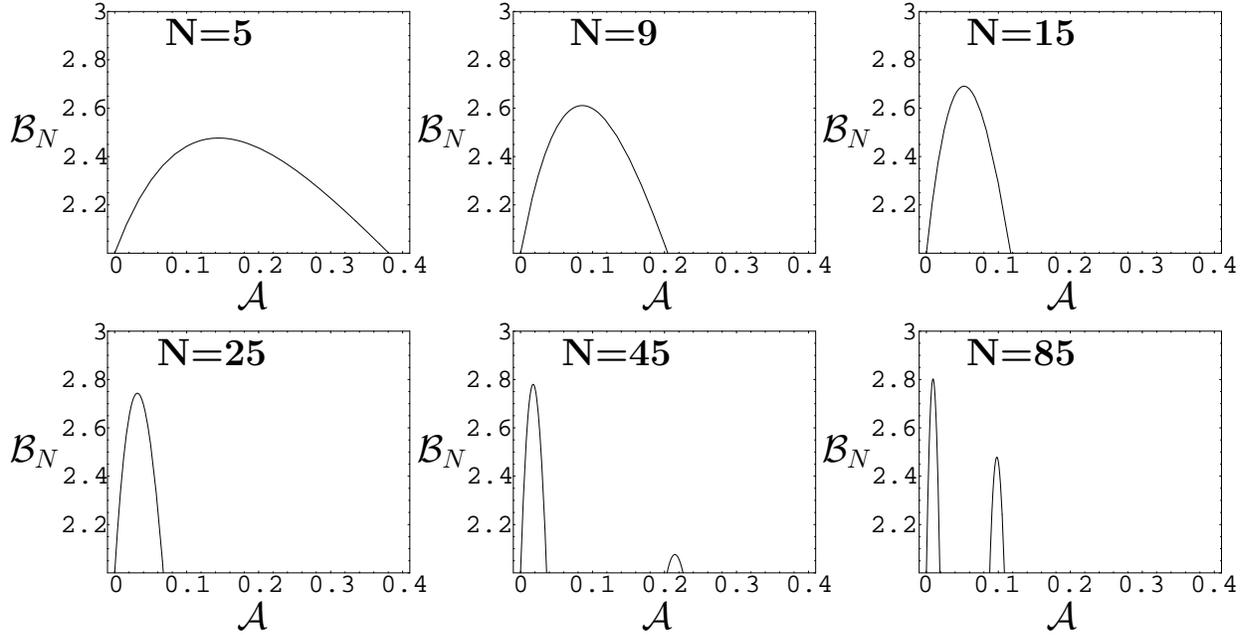}
\end{psfrags}
\end{center}
\caption{Maximum violations of the inequality $|{\mathcal{B}}_N|\leq 2$
imposed by local realistic theories in the limit of large squeezing.
${\mathcal{B}}_N$ is plotted as a function
of ${\mathcal{A}}\equiv {\mathcal{J}}e^{+2r}$ for different $N$.} 
\label{fig2}
\end{figure}

\begin{multicols}{2}
\noindent $\cdots=\alpha_{2k+1}'=i\sqrt{\mathcal{J}}$, the
remaining ones are
$\alpha_{2k+2}=\alpha_{2k+3}=\cdots=\alpha_N=0$, and $M=0,1,2,3,...$.
Because of the symmetry of the states $\Pi({\boldsymbol{\alpha}})$
in Eq.~(\ref{Wigneralpha}), all possible permutations of the 
($2k+1$) $\alpha_i'$'s with $\alpha_i'=i\sqrt{\mathcal{J}}$ and the 
[$N-(2k+1)$] $\alpha_i$'s with $\alpha_i=0$ can be described by
the same function $\Pi(\alpha_1',\alpha_2',...,\alpha_{2k+1}',
\alpha_{2k+2},\alpha_{2k+3},...,\alpha_N)$.

Similarly, with the same settings $\alpha_i'=i\sqrt{\mathcal{J}}$ and
$\alpha_i=0$, and again by exploiting symmetry, we obtain
\begin{eqnarray}\label{BN2}
{\rm for}\;N&=&5+8M:\;\;{\mathcal{B}}_N=2^{\frac{3-N}{2}}
\sum_{k=0}^{\frac{N-1}{2}}(-1)^{k+1} {N \choose 2k}\nonumber\\
&&\times\;\Pi(\alpha_1',\alpha_2',...,\alpha_{2k}',
\alpha_{2k+1},\alpha_{2k+2},...,\alpha_N),\\
\label{BN3}
{\rm for}\;N&=&7+8M:\;\;{\mathcal{B}}_N=2^{\frac{3-N}{2}}
\sum_{k=0}^{\frac{N-1}{2}}(-1)^{k+1} {N \choose 2k+1}\nonumber\\
&&\times\;\Pi(\alpha_1',\alpha_2',...,\alpha_{2k+1}',
\alpha_{2k+2},\alpha_{2k+3},...,\alpha_N),\\
\label{BN4}
{\rm for}\;N&=&9+8M:\;\;{\mathcal{B}}_N=2^{\frac{3-N}{2}}
\sum_{k=0}^{\frac{N-1}{2}}(-1)^k {N \choose 2k}\nonumber\\
&&\times\;\Pi(\alpha_1',\alpha_2',...,\alpha_{2k}',
\alpha_{2k+1},\alpha_{2k+2},...,\alpha_N).
\end{eqnarray} 
The functions concerned in these formulas are explicitly given by
[see Eq.~(\ref{Wigneralpha})]
\end{multicols}
\noindent\rule{5cm}{.6pt}
\begin{eqnarray}\label{explic1}
\Pi(\alpha_1',\alpha_2',...,\alpha_{2k}',
\alpha_{2k+1},\alpha_{2k+2},...,\alpha_N)&=&
\exp\left\{-2{\mathcal{J}}\cosh 2r\;(2k)+
2{\mathcal{J}}\sinh 2r\left[2k-2\frac{(2k)^2}{N}\right]\right\},\\
\label{explic2}
\Pi(\alpha_1',\alpha_2',...,\alpha_{2k+1}',
\alpha_{2k+2},\alpha_{2k+3},...,\alpha_N)&=&
\exp\left\{-2{\mathcal{J}}\cosh 2r\;(2k+1)+
2{\mathcal{J}}\sinh 2r\left[2k+1-2\frac{(2k+1)^2}{N}\right]\right\}.
\end{eqnarray}
\hfill\rule{5cm}{.6pt}
\begin{multicols}{2}
Let us first consider the case of zero squeezing, $r=0$.
The sum from Eq.~(\ref{BN1}) becomes in this case
\begin{eqnarray}\label{zerosq}
{\mathcal{B}}_N(r=0)&=&2^{\frac{3-N}{2}}
(1+e^{-4{\mathcal{J}}})^{N/2}\nonumber\\
&&\times\;\sin[N\arctan(e^{-2{\mathcal{J}}})].  
\end{eqnarray}

As expected, without squeezing, no violations 
of the Bell-type inequalities are obtained for the unentangled,
separable $N$-mode states: we find ${\mathcal{B}}_N(r=0)=2$ if
${\mathcal{J}}=0$ for any $N=3+8M$ and $|{\mathcal{B}}_N(r=0)|<2$
if ${\mathcal{J}}>0$. In the limit $N\to\infty$, we obtain
${\mathcal{B}}_N(r=0)\to 0$ for any ${\mathcal{J}}>0$.
Similar expressions as in Eq.~(\ref{zerosq}) can be found
for ${\mathcal{B}}_N(r=0)$ in the other cases of odd $N$,
$N=5+8M$, $N=7+8M$, and $N=9+8M$, and in fact, no violations occur.
The inequality $|{\mathcal{B}}_N|\leq 2$
imposed by local realistic theories always remains satisfied
for zero squeezing.

On the other hand, inferring from the results for $N\leq 5$ parties,
the maximum violations of $|{\mathcal{B}}_N|\leq 2$ occur for large
squeezing. Let us again consider the limit of large
\end{multicols}

\begin{figure}[htb]
\begin{center}
\begin{psfrags}
     \psfrag{B}[cb]{\Large ${\mathcal{B}}_N~~~$}
     \psfrag{J}[c]{\Large ${\mathcal{J}}$}
     \psfrag{a}{\bf r=0.1}
     \psfrag{b}{\bf r=0.3}
     \psfrag{c}{\bf r=0.8}
     \psfrag{d}{\bf r=1.5}
     \psfrag{N=5}{\Large \bf N=5}
     \psfrag{N=9}{\Large \bf N=9}
     \psfrag{N=15}{\Large \bf N=15}
     \psfrag{N=25}{\Large \bf N=25}
     \psfrag{N=45}{\Large \bf N=45}
     \psfrag{N=85}{\Large \bf N=85}
\epsfxsize=6.5in
     \epsfbox[-20 15 450 220]{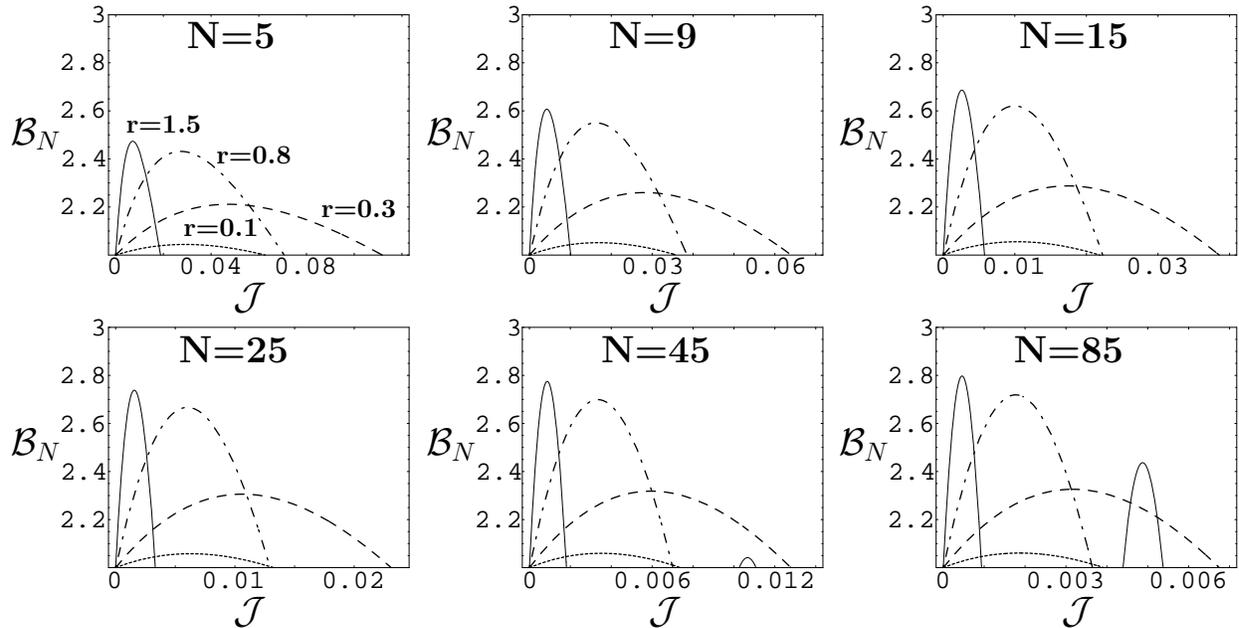}
\end{psfrags}
\end{center}
\caption{Violations of the inequality $|{\mathcal{B}}_N|\leq 2$
imposed by local realistic theories for different $N$ at certain
amounts of squeezing of the $N$-mode GHZ states:
$r=0.1$ ($\approx 0.9$ dB), $r=0.3$ ($\approx 2.6$ dB),
$r=0.8$ ($\approx 6.9$ dB), and $r=1.5$ ($\approx 13$ dB).
${\mathcal{B}}_N$ is plotted as a function of ${\mathcal{J}}$.
Note that the axes of the displacement parameter
${\mathcal{J}}$ vary in scale. The larger $N$
becomes, the smaller become the displacements required.} 
\label{fig3}
\end{figure}

\begin{multicols}{2}
\noindent squeezing ($\cosh 2r\approx\sinh 2r\approx e^{+2r}/2$) and define
${\mathcal{A}}\equiv {\mathcal{J}}e^{+2r}$. Now we can write 
Eq.~(\ref{explic1}) and Eq.~(\ref{explic2}) as
\begin{eqnarray}\label{explic1largesq}
\Pi(\alpha_1',\alpha_2',...,\alpha_{2k}',
\alpha_{2k+1},\alpha_{2k+2},...,\alpha_N)=\nonumber\\
\exp\left[-2{\mathcal{A}}\;(2k)^2/N\right],\\
\label{explic2largesq}
\Pi(\alpha_1',\alpha_2',...,\alpha_{2k+1}',
\alpha_{2k+2},\alpha_{2k+3},...,\alpha_N)=\nonumber\\
\exp\left[-2{\mathcal{A}}\;(2k+1)^2/N\right].
\end{eqnarray}
Figure 2 shows the maxima of the violations of $|{\mathcal{B}}_N|\leq 2$
(for our particular choice of settings), calculated with
Eq.~(\ref{BN1})-(\ref{BN4}) and the asymptotic results from
Eq.~(\ref{explic1largesq})-(\ref{explic2largesq}) for large squeezing.
The maximum violation grows from ${\mathcal{B}}_5^{\rm max}\approx 2.48$
for $N=5$ to ${\mathcal{B}}_{85}^{\rm max}\approx 2.8$ for $N=85$.
Within this range, a maximum violation near $2.8$ is already attained
with $N=45$ parties and there is only a very small increase from
$N=45$ to $N=85$. On the other hand, between $N=5$ and $N=9$, the maximum
violation goes up from $2.48$ to about $2.6$ which is still 
significantly less than the increase between $N=2$ 
(${\mathcal{B}}_2^{\rm max}\approx 2.19$) and $N=5$. This confirms our
conjecture based on the results for $N\leq 5$:
apparently, the maximum violation indeed grows with increasing number
of parties, but this growth seems to continuously decrease for larger
numbers of parties. In fact, from $N=45$ to $N=85$, we see a second
local maximum emerging rather than a significant further increase
of the absolute maximum violation.

In Fig.~3, calculated with Eq.~(\ref{BN1})-(\ref{BN4}) and 
Eq.~(\ref{explic1})-(\ref{explic2}),
violations of $|{\mathcal{B}}_N|\leq 2$ are compared between
different numbers of parties at certain amounts of squeezing of the
corresponding GHZ states. As stated earlier, the violations grow
with $N$ also for modest finite squeezing, but this increase is smaller
than the increase of the maximum violations and becomes unrecognizable
for small squeezing.
An illustrating example is that a violation comparable to the maximum
violation with the 2-mode EPR state for large squeezing
(${\mathcal{B}}_2^{\rm max}\approx 2.19$) can be attained with a
5-mode GHZ state built from five modestly squeezed states 
(about $2.6$ dB each).

We conclude with a summary and an assessment of our results.
We have considered pure multipartite entangled states described
by continuous quantum variables and shown that they violate Bell-type
inequalities imposed by local realism. An experimental
nonlocality test based on these states and on our scheme is
possible, but it would require detectors capable of resolving the 
number of absorbed photons \cite{Bana2}. Nevertheless, 
{\it the $N$-mode states which we have unambigously proven to exhibit
nonlocality can be relatively easily generated in practice, as opposed to 
the discrete-variable GHZ states on which all current
multiparty nonlocality proofs rely.}
Furthermore, entangled $N$-mode states similar to those considered here
can even be produced using only one single-mode squeezed vacuum state and 
linear optics instead of $N$ squeezed states \cite{PvL}. 
Since it has been shown
already that the entangled two-mode state created this way is nonlocal
with respect to parity measurements \cite{Wu}, one can apply our
analysis to the corresponding $N$-mode states and expect that they too
are nonlocal.

The degree of nonlocality of the continuous-variable GHZ states,
if represented by the maximum violation of the corresponding
Bell-type inequalities, seems to grow with increasing number of parties.
This growth, however, continuously decreases for larger numbers of 
parties. Thus, the evolution of the continuous-variable states' 
nonlocality with increasing number of parties and the corresponding
evolution of nonlocality for the qubit GHZ states are 
qualitatively equal but quantitatively different
(with an exponential increase for the qubits).
The reason for this may be that the latter always relies on 
maximally entangled states,
whereas the former depends on nonmaximally entangled states
as long as the squeezing remains finite. In fact, an observation
of the nonlocality of the continuous-variable states requires small
but nonzero displacements ${\mathcal{J}}\propto e^{-2r}$,
which is not achievable when the singular maximally entangled states 
for infinite squeezing are considered. 
 
This research was funded by a DAAD Doktorandenstipendium (HSP III) 
and by the EPSRC Grant No.\ GR/L91344.
PvL thanks T.C.Ralph, W.J.Munro, and A.K.Pati for
useful comments on the present paper and general discussions
concerning nonlocality.

\end{multicols}
\end{document}